\documentclass[aps,prl,reprint,floatfix,superscriptaddress,nofootinbib]{revtex4-2}

\usepackage{amsmath,amssymb,mathtools}
\usepackage{graphicx}
\usepackage{booktabs}
\usepackage{microtype}
\usepackage[hidelinks]{hyperref}

\makeatletter
\let\SM@addcontentsline\addcontentsline
\renewcommand{\addcontentsline}[3]{}
\newcommand{\SMtocstart}{\let\addcontentsline\SM@addcontentsline}
\makeatother

\begin{document}

\title{Frustration-Induced $1/6$-Vortex-Paired Phase in a Kagome XY Model}

\author{Tong-Yu Lin}
\affiliation{Department of Physics, Tsinghua University, Beijing 100084, China}

\author{Guang-Ming Zhang}
\email{zhanggm@shanghaitech.edu.cn}
\affiliation{State Key Laboratory of Quantum Functional Materials and School
of Physical Science and Technology, ShanghaiTech University, Shanghai 201210, China}
\affiliation{Department of Physics, Tsinghua University, Beijing 100084, China}

\date{\today}

\begin{abstract}
Geometric frustration can fractionalize vortices in two-dimensional XY systems, but a given frustrated model usually hosts only a single type of fractional vortex. Here we study a kagome XY model in which an antiferromagnetic coupling $J_1$ and a second-harmonic coupling $K$ favor two competing frustrated ground-state manifolds. For $K/J_1<1/9$, the ground states map onto those of the kagome three-state Potts antiferromagnet and support $1/3$ vortices; for $K/J_1>1/9$, they map onto those of the kagome Ising antiferromagnet and support $1/2$ vortices. At $K/J_1=1/9$, the two manifolds become degenerate, and bends in the domain walls between them carry $1/6$-vortex excitations. Tensor-network calculations show that a finite-temperature phase with quasi-long-range order only in $e^{6i\theta}$ emerges from this degenerate point, separating the $1/3$- and $1/2$-vortex-paired phases. This $1/6$-vortex-paired phase has no counterpart in either parent manifold, showing that competing frustrated manifolds provide a route to new fractional-vortex-paired phases.
\end{abstract}

\maketitle

\noindent\textbf{Introduction---} Frustration arises when local interactions
cannot be minimized simultaneously. In classical spin systems it often leads to
additional ground-state degeneracy, unconventional excitations, and
finite-temperature behavior absent in unfrustrated
systems~\cite{Moessner_2001,Ramirez_1994,Lacroix_2011,Diep_2020}.
Two-dimensional XY systems offer a particularly rich setting for these effects,
because frustration acts not only on the ground states but also on the
topological defects. In the unfrustrated XY model,
the elementary defects are integer vortices. At low temperatures, vortices and
antivortices remain bound in pairs, and the phase field $e^{i\theta}$ exhibits
quasi-long-range order (QLRO); this low-temperature phase can thus be regarded
as an integer-vortex-paired phase. Upon further heating, the system undergoes a
Kosterlitz--Thouless (KT) transition driven by the dissociation of integer
vortex--antivortex pairs \cite{Berezinsky1971,KosterlitzThouless1973}.

Frustration can instead permit fractional vortices carrying a rational
topological charge $1/n$, and even fractional-vortex-paired
phases~\cite{Korshunov_1986,Korshunov1986General,Korshunov_2006}. When
frustration is strong enough to produce an extensive ground-state
degeneracy~\cite{Korshunov2002Kagome,Korshunov2005NoVortexOrder}, the order in
$e^{i\theta}$ is suppressed in the ground-state manifold, which is
instead characterized by order in a composite field $e^{in\theta}$. The
relevant topological excitations are then $1/n$ vortices rather than integer
vortices. Their origin lies in the additional ground-state degeneracy, which
can be described by discrete degrees of freedom: domain walls can be
constructed between different ground states, and fractional vortices form at
bends of these walls~\cite{Halsey1985,Korshunov2002Kagome}. The low-temperature
phase, in which these $1/n$ vortices remain bound in pairs and $e^{in\theta}$
exhibits QLRO, is consequently referred to as a $1/n$-vortex-paired
phase~\cite{Korshunov2002Kagome,Korshunov2012Honeycomb,Korshunov2005Dice}.
These phases are especially interesting in
superconducting systems. If the XY degree of freedom describes charge-$2e$
superconducting fluctuations, then $e^{in\theta}$ represents charge-$2ne$
superconducting order
\cite{Berg2009Charge4e,Agterberg2011Charge6,Agterberg2020PDW,
Pan2022Charge6e,ZhouWang2022Charge46e,Liu2023Charge4e,
LinSongZhang2025Charge6e,VarmaWang2023Flux,
Fernandes2021Charge4e,Jian2021Charge4e,Song2022HalfVortex,
Hecker2023Charge4e,Hecker2024DomainWall4e,
Weerda2026Dice4e,Wang2025Roton6e,Huecker2026Charge4e,
Zou2026Charge4e,Gao2026Charge46e}, while a $1/n$ vortex carries the
reduced magnetic flux quantum $\Phi_0/n=h/(2ne)$, detectable by SQUID
magnetometry \cite{Iguchi2023FractionalFlux} or magnetoresistance
oscillations \cite{Ge2024Flux}.

In frustrated XY models, the fractional vortices are thus set by the
ground-state manifold. A given model, however, typically hosts only a single
extensively degenerate manifold, and therefore only one type of fractional
vortex~\cite{Korshunov2005NoVortexOrder,Korshunov2012Honeycomb,
Korshunov2002Kagome}. This raises the central question of this work: in a frustrated
XY model with several distinct extensively degenerate ground-state manifolds,
can multiple fractional-vortex-paired phases form, and can the competition
between them give rise to fractional vortices and phases absent from any single
manifold?

To address this question, we start from the antiferromagnetic kagome XY model
with nearest-neighbor coupling $J_1$, whose extensively degenerate ground
states map onto those of the kagome antiferromagnetic three-state Potts model
and support $1/3$ vortices~\cite{HuseRutenberg1992,Korshunov2002Kagome,
Song_2023}. We then add a
second-harmonic coupling $K$. Higher-harmonic couplings have been studied
extensively in generalized XY
models~\cite{LeeGrinstein1985,Korshunov1986Modified,CarpenterChalker1989,
ShiLamacraftFendley2011,Poderoso2011,Canova2016}. There, the ground state is the
ferromagnetic XY state without additional degeneracy, the $1/q$ vortices are
inherited directly from the $2\pi/q$ periodicity of the added $\cos(q\theta)$
coupling, and the $1/q$-vortex-paired phase forms only where this coupling
dominates~\cite{LeeGrinstein1985,Poderoso2011}. The role of $K$ here is
entirely different. The frustration of $J_1$ remains dominant, and $K$ instead
changes the structure of the frustrated ground states, converting them into a
second extensively degenerate manifold that maps onto the kagome Ising
antiferromagnet and supports $1/2$ vortices.

The combination of frustration with higher-harmonic
couplings~\cite{Park2008,LachZukovic2020,LachZukovic2021}, as well as the
competition between different higher-harmonic
couplings~\cite{ZukovicKalagov2018}, has been explored previously and leads to
a variety of new phases. However, the possibility that a higher-harmonic
coupling generates a new extensively degenerate manifold, and correspondingly a
new fractional-vortex-paired phase, has received little attention. In the
present model, we find that the two frustrated ground-state manifolds become
degenerate at $K/J_1=1/9$, forming a new extensively degenerate ground-state
manifold, and that bends in the domain walls between them carry $1/6$
vortices. Combining this analysis with tensor-network calculations, we further
show that a $1/6$-vortex-paired phase forms in the vicinity of this point.

\begin{figure}[t]
  \centering
  \includegraphics[width=\columnwidth]{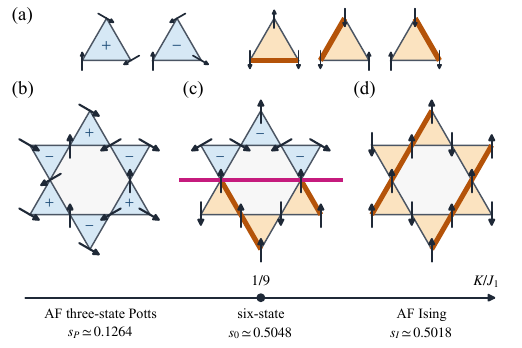}
  \caption{Ground states of the kagome $J_1$--$K$ XY model. (a) The two $P$-type
  (chirality $\pm$) and three $I$-type triangle configurations; the thick bond
  connects the two parallel spins of an $I$ triangle. (b)--(d) Representative
  ground states for $K/J_1<1/9$, $=1/9$, and $>1/9$; the magenta line in (c)
  represents a straight $P$--$I$ domain wall.}
  \label{fig:groundstate}
\end{figure}

\noindent\textbf{Model and ground states---} We consider the frustrated $J_1$--$K$ XY model on the kagome lattice,
\begin{equation}
  \mathcal{H}=\sum_{\langle ij\rangle}
  \left[J_1\cos(\theta_i-\theta_j)-K\cos2(\theta_i-\theta_j)\right],
  \label{eq:hamiltonian}
\end{equation}
where $J_1,K>0$ act on nearest-neighbor bonds. Since each bond belongs to
exactly one triangle, the ground states are obtained by minimizing the energy
of every triangle simultaneously.
As the energy depends only on angle differences, we describe ground states
by the bond phase differences
$\phi_{ij}=\operatorname{mod}_{2\pi}(\theta_i-\theta_j)\in(-\pi,\pi]$, which
removes the global $U(1)$ degree of freedom.

For $K/J_1<1/9$, the triangle energy is minimized by the $120^\circ$
configuration, which we call the $P$ type. In this configuration every phase
difference on the triangle equals $\pm2\pi/3$, and a single $P$-type triangle
has two possible configurations, distinguished by their chirality $\pm$
[Fig.~\ref{fig:groundstate}(a)]. Assembling these triangles into a lattice
ground state [Fig.~\ref{fig:groundstate}(b)] requires single-valued spin angles, so the
phase differences around every hexagon obey the compatibility condition
\begin{equation}
 \sum_{\langle ij\rangle\in\partial h}\phi_{ij}=2\pi n_h,
 \qquad n_h\in\mathbb Z.
 \label{eq:hexagon_integer_winding}
\end{equation}
With only two allowed values on every bond, the
resulting ground-state manifold maps onto the kagome antiferromagnetic
three-state Potts model. The exact solution of that
model \cite{Baxter_1970} therefore implies that the $P$-type ground-state ensemble of the XY
model is extensively degenerate, with residual entropy $s_P\simeq0.1264$ per
spin. Within this ensemble, the composite field $e^{3i\theta}$ has long-range
order, whereas $e^{i\theta}$ exhibits only QLRO~\cite{HuseRutenberg1992}.

For $K/J_1>1/9$, the triangle energy is instead minimized by the collinear
configuration with two parallel spins and one antiparallel spin, which we call
the $I$ type. Its phase differences take the values $0$ and $\pi$, and a single
$I$-type triangle has three possible configurations, corresponding to the
position of the two parallel spins [Fig.~\ref{fig:groundstate}(a)]. Ground states assembled from these triangles
[Fig.~\ref{fig:groundstate}(d)] again obey
Eq.~(\ref{eq:hexagon_integer_winding}). This ground-state manifold maps
onto the kagome antiferromagnetic Ising model, and its exact solution
\cite{Kano_1953} implies that the $I$-type ground-state ensemble
of the XY model is extensively degenerate, with residual entropy
$s_I\simeq0.50183$ per spin. Here $e^{2i\theta}$ has long-range order, whereas
the correlations of $e^{i\theta}$ are short-ranged.

At $K/J_1=1/9$, the local $P$- and $I$-type triangle configurations become
degenerate. Both types of triangles can coexist in a ground-state
configuration, as illustrated in Fig.~\ref{fig:groundstate}(c), where $P$ and
$I$ triangles meet along a straight $P$--$I$ domain wall. Such a mixed configuration is a ground state
provided that the phase differences still satisfy the compatibility condition
in Eq.~(\ref{eq:hexagon_integer_winding}) on every hexagon. Since the allowed
phase differences now include $\pm2\pi/3$ from the $P$ triangles and $0$ or
$\pi$ from the $I$ triangles, fixing a reference angle $\Theta$ restricts the
spins to six orientations, $\theta_i=\Theta+\pi a_i/3$ with
$a_i\in\mathbb Z_6$. The resulting ground-state manifold can therefore be represented as a
constrained six-state spin model, which directly implies long-range order in
$e^{6i\theta}$. Our tensor-network calculation gives a residual
entropy $s_0\simeq0.50476$ per spin and confirms that the ground-state ensemble
also retains long-range order in $e^{2i\theta}$. Details of this
calculation are given in the Supplemental Material.

\begin{figure}[t]
  \centering
  \includegraphics[width=\columnwidth]{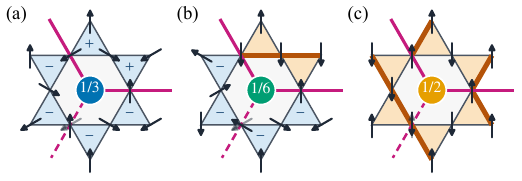}
  \caption{(a) $1/3$, (b) $1/6$, and (c) $1/2$ vortices at a $120^\circ$ bend of a
  $P$--$P$, $P$--$I$, and $I$--$I$ domain wall (solid line). The dashed line is
  the branch cut; the spin mismatch on the site it crosses is shown by the
  translucent arrow.}
  \label{fig:fractional_vortices}
\end{figure}

\noindent\textbf{Fractional-vortex excitations---} Having established the
ground-state properties, we next consider the topological excitations and the
finite-temperature phases that they can produce. The $P$ sector for
$K/J_1<1/9$ has the same ground-state constraint as the $K=0$ frustrated
antiferromagnetic XY model, so the construction of fractional vortices in that
model can be used directly here \cite{Korshunov2002Kagome}.

The excitations can be characterized by studying domain-wall structures in the
ground-state manifold. We first choose a uniform ground-state domain as a
reference, in which every triangle has the same phase-difference pattern.
Different uniform ground-state domains can then be joined by a domain wall. If
inserting the wall preserves the hexagonal compatibility condition on every
hexagon, it costs no additional energy; a straight $P$--$P$ domain wall is one
such zero-energy example. When the wall has a $120^\circ$ bend, as in
Fig.~\ref{fig:fractional_vortices}(a), the phase-difference patterns on the two
sides of the wall correspond to different uniform domains. Continuing the
ground state around the hexagon at the bend then fails to return to the
starting spin: the two values on the site crossed by the branch cut differ by
a phase mismatch $\delta\theta$, shown by the translucent arrow in
Fig.~\ref{fig:fractional_vortices}.
The continuous spin angles relax to compensate this mismatch, producing an
effective fractional vortex centered on that hexagon, with fractional winding
\begin{equation}
 \frac{1}{n}=\frac{\delta\theta}{2\pi}.
 \label{eq:fractional_winding_mismatch}
\end{equation}
For the $P$-type reference state, the bend of a $P$--$P$ wall gives
$\delta\theta=\pm2\pi/3$ and hence a $1/3$ vortex
[Fig.~\ref{fig:fractional_vortices}(a)]. The same construction applies on the
$K/J_1>1/9$ side using a uniform $I$-type reference state. A bend of an
$I$--$I$ domain wall produces $\delta\theta=\pi$ and therefore a $1/2$ vortex
[Fig.~\ref{fig:fractional_vortices}(c)].

At $K/J_1=1/9$, the $P$ and $I$ manifolds are degenerate, so straight
$P$--$I$ domain walls also cost no additional energy. A bend of such a wall can produce the
smaller mismatch $\delta\theta=\pm\pi/3$, yielding a $1/6$ vortex
[Fig.~\ref{fig:fractional_vortices}(b)]. Thus, the coexistence of the two degenerate
ground-state sectors permits a fractional-vortex excitation that is absent
from either pure sector.

Because a fractional vortex is attached to a domain wall, its excitation
energy takes the form
\begin{equation}
 E_{\rm fv}(L)=E_{\rm dw}L+\frac{E_{\rm v}}{n^2}\ln L,
 \label{eq:fractional_vortex_energy}
\end{equation}
where $E_{\rm dw}$ is the domain-wall energy per unit length, $E_{\rm v}\ln L$
is the energy of an integer vortex, $1/n$ is the fractional winding in
Eq.~(\ref{eq:fractional_winding_mismatch}), and $L$ is the system size. Away
from $K/J_1=1/9$, only the domain walls within the corresponding parent manifold
have zero line tension: $P$--$P$ walls on the $K/J_1<1/9$ side and $I$--$I$
walls on the $K/J_1>1/9$ side. The lowest-energy fractional excitations are
therefore $1/3$ and $1/2$ vortices, respectively, suggesting the associated
$1/3$- and $1/2$-vortex-paired phases at low temperature. At
$K/J_1=1/9$, mixed $P$--$I$ walls also have zero energy, making the $1/6$
vortex the lowest-energy fractional excitation. Based on the ground-state and
excitation analysis, in the finite-temperature region near $K/J_1=1/9$,
the proliferation of $P$--$I$ domain walls mixes the $P$- and $I$-type
ground-state manifolds. The resulting
low-temperature phase is governed by QLRO in $e^{6i\theta}$, with the
$1/6$ vortex as its relevant topological defect.

\noindent\textbf{Tensor network formulation---} To further test the physical
picture proposed above, we study the finite-temperature properties of the model
by directly evaluating its partition function with the tensor-network method
developed for frustrated classical spin systems \cite{Song2023}. The partition
function is represented as a contraction of local tensors constructed from the
Boltzmann weight of an elementary kagome triangle. After Fourier transformation
of the angular variables and regrouping of the local tensors, the resulting
network can be viewed as a uniform tensor network, whose row-to-row contraction
defines an infinite transfer operator $\mathbb{T}$. The dominant eigenvalue and
eigenvector are obtained from the maximal-eigenvalue equation
$\mathbb{T}|\Psi(A)\rangle=\Lambda_{\max}|\Psi(A)\rangle$
using the VUMPS algorithm \cite{ZaunerStauber2018,Vanderstraeten_2019,Fishman_2018}.
Thermodynamic quantities, spin-spin correlation functions, and the
entanglement entropy are evaluated from the dominant transfer-matrix fixed
point and corresponding impurity tensors. The Fourier cutoff $n$ and the MPS bond
dimension $D$ control the numerical accuracy; all data reported below are obtained
after increasing these dimensions until the observables become stable. Further
details of the tensor construction and numerical implementation are given in
the Supplemental Material.

\begin{figure}[t]
  \centering
  \includegraphics[width=\columnwidth]{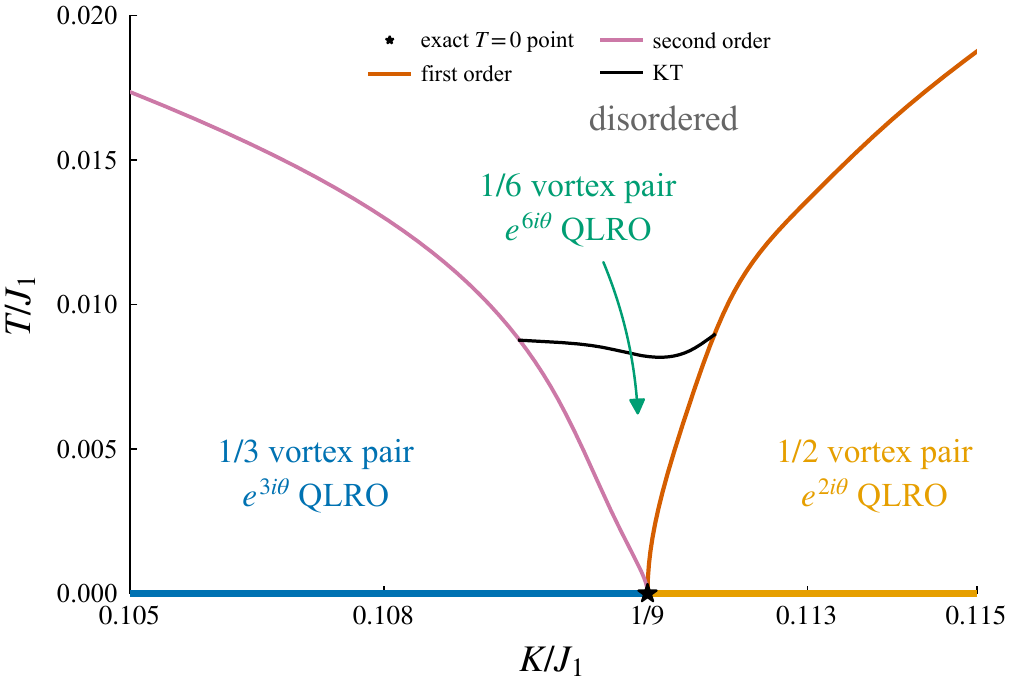}
  \caption{Finite-temperature phase diagram of the kagome $J_1$--$K$ XY model.
  The phase boundaries are interpolated through the transition temperatures
  obtained from the entanglement-entropy peaks; line colors indicate first-order,
  second-order, and KT transitions.}
  \label{fig:finite_temperature_phase_diagram}
\end{figure}

\noindent\textbf{Finite-temperature phase diagram---} With this tensor-network setup, we calculate the finite-temperature partition
function and verify the fractional-vortex picture above. The resulting phase
diagram is shown in Fig.~\ref{fig:finite_temperature_phase_diagram}.

\begin{figure}[!t]
  \centering
  \includegraphics[width=\columnwidth]{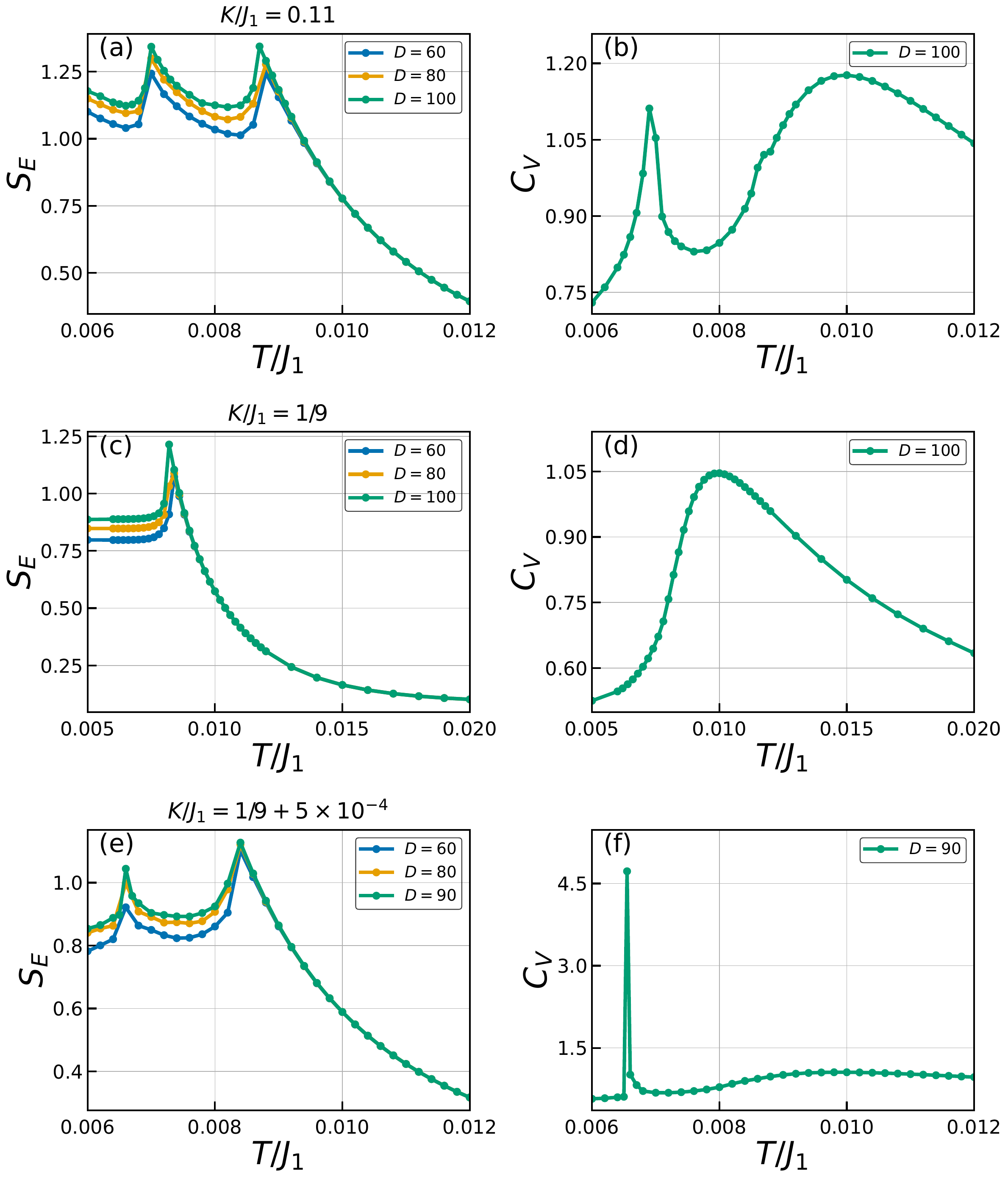}
  \caption{Entanglement entropy $S_E$ and specific heat $C_V$ for
  representative temperature scans at (a),(b) $K/J_1=0.11$, (c),(d) $1/9$, and
  (e),(f) $1/9+5\times10^{-4}$.}
  \label{fig:thermal_diagnostics}
\end{figure}

We determine the phase boundaries from singularities in the entanglement
entropy of the dominant transfer-matrix eigenvector and use the specific heat
to characterize the transitions. Figure~\ref{fig:thermal_diagnostics}
illustrates this procedure through three representative temperature scans near
$K/J_1=1/9$. At $K/J_1=1/9$ itself, the entanglement entropy exhibits a
single peak [Fig.~\ref{fig:thermal_diagnostics}(c)], while the specific heat
shows only a broad, rounded maximum [Fig.~\ref{fig:thermal_diagnostics}(d)],
consistent with a KT transition to the disordered phase. At
$K/J_1=0.11<1/9$, two entropy peaks resolve two successive transitions
[Fig.~\ref{fig:thermal_diagnostics}(a)]. The lower-temperature transition is
accompanied by a sharp specific-heat anomaly, consistent with a second-order
transition, whereas the specific heat remains smooth across the
higher-temperature transition, consistent with the KT type
[Fig.~\ref{fig:thermal_diagnostics}(b)]. At $K/J_1=1/9+5\times10^{-4}$, the
entropy likewise exhibits two peaks [Fig.~\ref{fig:thermal_diagnostics}(e)].
The lower-temperature transition is now accompanied by a jump in the internal
energy, which produces a divergence in the specific heat, consistent with a
first-order transition, while the upper one remains consistent with the KT type
[Fig.~\ref{fig:thermal_diagnostics}(f)]. Repeating these temperature
scans at different $K/J_1$ traces the phase boundaries shown in
Fig.~\ref{fig:finite_temperature_phase_diagram}. As
$K/J_1$ moves farther from $1/9$ on either side, the two entropy peaks merge
into a single peak, indicating that the intermediate-temperature phase
disappears and the low-temperature phase undergoes a direct transition to the
disordered phase. These results delineate an intermediate phase extending to
both sides of $K/J_1=1/9$.

\begin{figure}[t]
  \centering
  \includegraphics[width=\columnwidth]{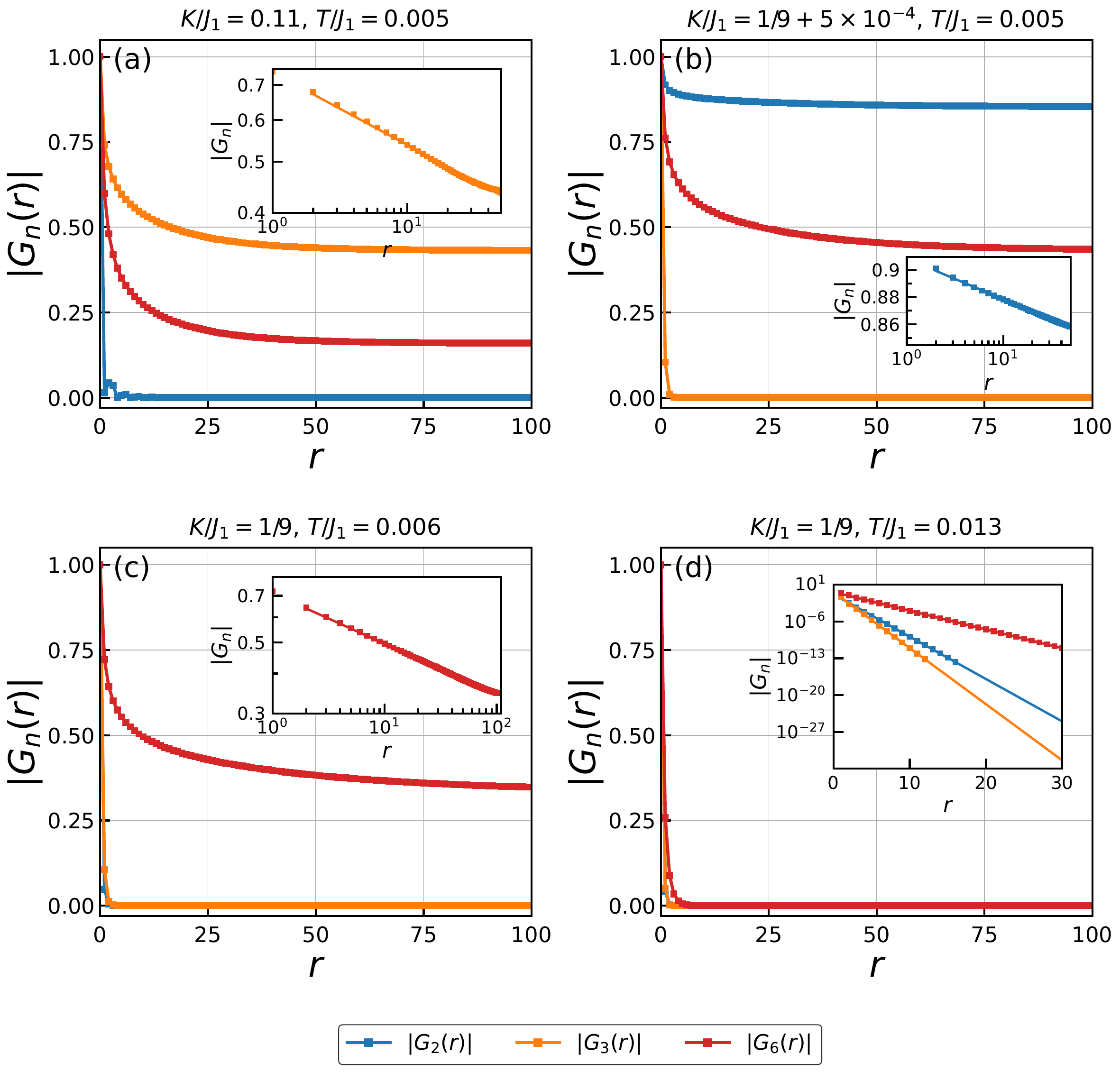}
  \caption{Spin-spin correlation functions $|G_n(r)|$ for $n=2,3,6$ in the
  $1/3$-vortex-paired phase (a), the $1/2$-vortex-paired phase (b), the
  intermediate $1/6$-vortex-paired phase at $K/J_1=1/9$ (c), and the
  high-temperature disordered phase (d). The insets in (a)--(c) show $|G_3|$,
  $|G_2|$, and $|G_6|$, respectively, on log-log scales with power-law fits
  $|G_n(r)|\sim r^{-\eta_n}$. The inset in (d) shows all three on a
  semilogarithmic scale with exponential fits $|G_n(r)|\sim e^{-r/\xi_n}$.}
  \label{fig:G6_low_high_decay}
\end{figure}

Having determined the phase boundaries, we identify the nature of each phase
from the spin-spin correlation functions
$G_n(r)=\langle e^{in[\theta(r)-\theta(0)]}\rangle$, where $n$ is an integer
and $r$ is the separation between the two spins. Each vortex-paired
phase is identified by the smallest $n$ for which $e^{in\theta}$ exhibits
QLRO. Representative data points illustrating the correlations in each phase
are shown in Fig.~\ref{fig:G6_low_high_decay}. In the low-temperature phase on the
$K/J_1<1/9$ side,
$G_3(r)$ and $G_6(r)$ decay algebraically, whereas $G_2(r)$ is short-ranged
[Fig.~\ref{fig:G6_low_high_decay}(a)], identifying the $1/3$-vortex-paired
phase. On the $K/J_1>1/9$ side, the low-temperature phase instead exhibits
algebraic decay in $G_2(r)$ and $G_6(r)$, while $G_3(r)$ is short-ranged
[Fig.~\ref{fig:G6_low_high_decay}(b)], identifying the $1/2$-vortex-paired
phase. The lowest-order phase coherence in these two phases is thus
consistent with the ground-state LRO characterized by
$e^{3i\theta}$ and $e^{2i\theta}$ on the
$K/J_1<1/9$ and $K/J_1>1/9$ sides, respectively.

In the intermediate phase, represented by the low-temperature point at
$K/J_1=1/9$ in Fig.~\ref{fig:G6_low_high_decay}(c), both $G_2(r)$ and
$G_3(r)$ are short-ranged, while $G_6(r)$ retains algebraic decay, so
$e^{6i\theta}$ is the lowest-order field with QLRO, identifying the
intermediate phase as a $1/6$-vortex-paired phase.
Upon further heating from the intermediate phase into the disordered
phase, $G_6(r)$ changes from algebraic to exponential decay
[Fig.~\ref{fig:G6_low_high_decay}(d)], consistent with a transition
driven by the unbinding of $1/6$ vortex--antivortex pairs.

\noindent\textbf{Discussion and conclusion---} In summary, we have shown that
in the kagome $J_1$--$K$ XY model frustration produces two extensively
degenerate ground-state manifolds, which support $1/3$- and
$1/2$-vortex-paired phases, respectively. To our knowledge, this provides the
first realization of multiple frustration-induced fractional-vortex-paired
phases in a single lattice model. The two manifolds become degenerate at
$K/J_1=1/9$, from which a $1/6$-vortex-paired phase emerges that has no
counterpart in either parent manifold. Whether the competition between other
frustrated ground-state manifolds can generate further fractional-vortex-paired
phases remains an open question.

Several aspects of the phase diagram deserve further study. At
$K/J_1=1/9$, the equal-weight $T=0$ ground-state ensemble has long-range
order in $e^{2i\theta}$, whereas the $e^{6i\theta}$ phase in
Fig.~\ref{fig:finite_temperature_phase_diagram} emerges directly from this point at finite temperature. This behavior
can be understood by considering fluctuations around the degenerate ground
states, which reweight the ground-state ensemble through a mechanism known
as order by disorder
\cite{Villain1980Order,MoessnerChalker1998PRL,MoessnerChalker1998PRB}.
The analysis in the Supplemental Material suggests that, as $T\to0^+$,
the order in $e^{2i\theta}$ disappears, leaving $e^{6i\theta}$ as the
lowest-order field that retains order.

The behavior away from $K/J_1=1/9$ requires a more
detailed analysis. The transitions from the $e^{2i\theta}$ and $e^{3i\theta}$
phases to the disordered phase appear to be continuous or first order in the
parameter ranges studied here, whereas the transitions farther from
$K/J_1=1/9$ are expected to be of the KT type. How these
different transitions evolve into one another remains to be clarified.
Including further-neighbor interactions would provide another interesting
direction. In the antiferromagnetic kagome Ising model, second- and
third-neighbor interactions produce a rich phase diagram
\cite{Vanhecke_2021,Colbois2021Artificial,Colbois_2022}; combining such
interactions with the XY degrees of freedom may lead to additional phases.

Finally, we briefly discuss possible experimental realizations of the
model. Josephson-junction arrays provide a promising platform: in a kagome
array with first- and second-harmonic Josephson couplings, an applied magnetic
field introducing half a flux quantum through each triangular plaquette can
produce an effective antiferromagnetic $J_1$ coupling while preserving the
ferromagnetic sign of $K$. Kagome-lattice superconductors offer another
route. In our previous work, we showed that the antiferromagnetic $J_1$--$J_2$
XY model serves as an effective theory for the melting of the commensurate
$3Q$ pair-density-wave (PDW) state on the kagome lattice, giving rise to a
charge-$6e$ condensed phase \cite{LinSongZhang2025Charge6e}. Building on this
effective description, it would be interesting to explore how the
second-harmonic coupling $K$ can arise in such PDW systems, and thus how the
present model can be realized in specific kagome materials.

\begin{acknowledgments}
The authors are very grateful to Xia-Ze Xu for stimulating discussions. This
research is supported by the National Key Research and Development Program of
China (Grant No.~2023YFA1406400).
\end{acknowledgments}

\bibliography{references}

\clearpage
\onecolumngrid
\setcounter{equation}{0}
\setcounter{figure}{0}
\setcounter{table}{0}
\setcounter{section}{0}
\setcounter{secnumdepth}{2}
\setcounter{tocdepth}{2}
\renewcommand{\theequation}{S\arabic{equation}}
\renewcommand{\thefigure}{S\arabic{figure}}
\renewcommand{\thetable}{S\arabic{table}}
\renewcommand{\thesection}{S\arabic{section}}

\begin{center}
{\large\bfseries Supplemental Material for\\[2pt]
``Frustration-Induced $1/6$-Vortex-Paired Phase in a Kagome XY Model''}\\[10pt]
Tong-Yu Lin$^{1}$ and Guang-Ming Zhang$^{2,1}$\\[4pt]
{\small\itshape $^{1}$Department of Physics, Tsinghua University, Beijing 100084, China\\
$^{2}$State Key Laboratory of Quantum Functional Materials and School of Physical
Science and Technology,\\ ShanghaiTech University, Shanghai 201210, China}
\end{center}
\vspace{6pt}

\tableofcontents
\SMtocstart
\vspace{12pt}
\twocolumngrid

\section{Ground-state ensemble and its low-temperature reweighting}
\label{sec:ground}

\subsection{Local ground-state rules}

The Hamiltonian of the kagome $J_1$--$K$ XY model is
\begin{equation}
 \mathcal H=\sum_{\langle ij\rangle}
 \left[J_1\cos(\theta_i-\theta_j)-K\cos2(\theta_i-\theta_j)\right],
 \label{eq:S-hamiltonian}
\end{equation}
with $J_1>0$ and $K\geq0$. Each nearest-neighbor bond belongs to exactly one
elementary triangle, so $\mathcal H=\sum_t h_t$, where $t$ runs over up and
down triangles and
\begin{equation}
 \begin{split}
 h_t(\theta_1,\theta_2,\theta_3)
 ={}&\sum_{(ij)=(12),(23),(31)}
 \bigl[J_1\cos(\theta_i-\theta_j)\\
 &\hspace{34pt}-K\cos2(\theta_i-\theta_j)\bigr].
 \end{split}
 \label{eq:triangle-energy}
\end{equation}
Any configuration that minimizes every $h_t$ simultaneously is therefore a
global ground state.

Two local configurations are relevant: the $120^\circ$ configuration
$P:\ (\theta_1,\theta_2,\theta_3)=\Theta+(0,2\pi/3,4\pi/3)$ and the collinear
configuration $I:\ (\theta_1,\theta_2,\theta_3)=\Theta+(0,0,\pi)$, together
with their permutations. Their energies are
\begin{equation}
 h_P=-\frac32J_1+\frac32K,\qquad h_I=-J_1-3K,
 \label{eq:local-energies}
\end{equation}
so that $h_I-h_P=(J_1-9K)/2$. Ground states are thus built from $P$ triangles
for $K/J_1<1/9$, from $I$ triangles for $K/J_1>1/9$, and from both at the
degenerate point $K/J_1=1/9$. Up to a global rotation, an isolated $P$
triangle has two configurations and an $I$ triangle has three, corresponding
to the position of the two parallel spins.

Once a reference angle $\Theta$ is fixed, the spins of a $P$-type ground state
take three values separated by $2\pi/3$ and those of an $I$-type ground state
take two values separated by $\pi$. These constraints yield the mappings onto
the kagome antiferromagnetic three-state Potts and Ising models, respectively
\cite{Baxter_1970,Kano_1953}. At $K/J_1=1/9$, both $P$ and $I$ triangles
are allowed. Since their angle differences, $\pm2\pi/3$ and $0,\pi$, are all
multiples of $\pi/3$, the ground states map onto a six-state spin model with
\begin{equation}
 \theta_i=\Theta+\frac{\pi a_i}{3},\qquad a_i\in\{0,1,\dots,5\}.
 \label{eq:six-state}
\end{equation}

\begin{figure*}[t]
 \centering
 \includegraphics[width=0.76\textwidth]{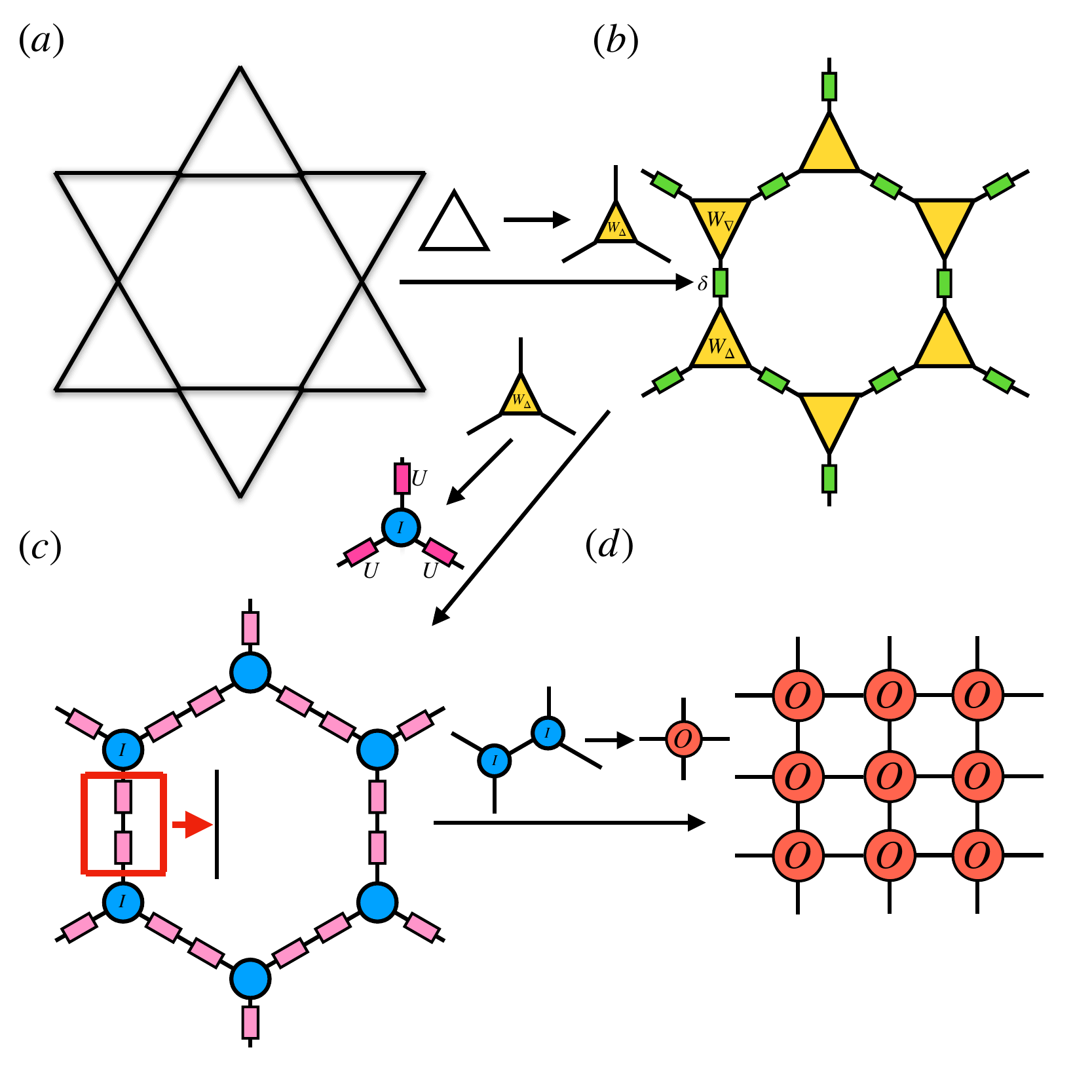}
 \caption{Triangle-based tensor-network representation of the kagome
 $J_1$--$K$ XY model. (a) Kagome lattice and its elementary triangles.
 (b) Local weights $W_\Delta$ and $W_\nabla$ on up and down triangles, with
 equality constraints $\delta$ on shared spins. For ground-state counting the
 variables are six-state labels and $W$ is the constraint tensor of
 Eq.~\eqref{eq:constraint}; at finite temperature they are the spin angles
 with Boltzmann weights. (c) A Fourier transformation $U$ yields discrete
 tensors $I$, and integration over a shared angle reduces the Fourier basis
 functions to an identity bond. (d) Contracting an up and a down triangle gives
 a rank-four tensor $O$ on a square lattice.}
 \label{fig:construction}
\end{figure*}

\subsection{Ground-state counting}

Since a global ground state is assembled from local triangle ground states
that agree on their shared spins, the ground-state manifold can be represented
and counted by a tensor network built from triangle tensors
\cite{Song2023}. Each local tensor encodes the allowed ground-state
configurations of a single triangle. As shown in
Fig.~\ref{fig:construction}(a) and \ref{fig:construction}(b), we place a
rank-three tensor $W_\Delta$ ($W_\nabla$) on each up (down) triangle, with one
leg for each of its three spins, labeled by the six-state variable of
Eq.~\eqref{eq:six-state}. A tensor element equals one if the three labels
form an allowed local ground state and zero otherwise. To write these elements
explicitly, let $\chi_P(a,b,c)=1$ if $(a,b,c)$ is a
permutation of $(a_0,a_0+2,a_0+4)$ modulo six for some $a_0$, and
$\chi_I(a,b,c)=1$ if it is a permutation of $(a_0,a_0,a_0+3)$; otherwise both
vanish. Both triangle orientations use the same tensor,
$W_\Delta(a,b,c)=W_\nabla(a,b,c)=W_{abc}$, with
\begin{equation}
 W_{abc}=\begin{cases}
 \chi_P(a,b,c),&K/J_1<1/9,\\
 \chi_P(a,b,c)+\chi_I(a,b,c),&K/J_1=1/9,\\
 \chi_I(a,b,c),&K/J_1>1/9.
 \end{cases}
 \label{eq:constraint}
\end{equation}
Away from $K/J_1=1/9$ the local space could be reduced to three or two states;
the six-state form treats all three cases uniformly.

With these tensors, the number of ground states is given by the
zero-temperature partition function
\begin{equation}
 Z_0=\sum_{\{a_i\}}\prod_t W_{a_tb_tc_t},
 \label{eq:counting}
\end{equation}
where $a_t,b_t,c_t$ are the labels of the three spins of triangle
$t$. A configuration $\{a_i\}$ contributes one if every triangle satisfies the
local rule and zero otherwise, so $Z_0$ counts the ground states for a fixed
reference angle $\Theta$. In the tensor network, a spin shared by two
neighboring triangles is represented by a Kronecker delta on the link between
them [Fig.~\ref{fig:construction}(b)]; absorbing these deltas into the $W$
tensors turns Eq.~\eqref{eq:counting} into the contraction of a honeycomb
network. Contracting the
shared index of an adjacent up--down pair gives
\begin{equation}
 O_{abcd}=\sum_{g=0}^{5} W_{abg}W_{gcd},
 \label{eq:ground-mpo}
\end{equation}
where $g$ labels the spin shared within the pair. The resulting square-lattice
network [Fig.~\ref{fig:construction}(d)] has one cell per two triangles, i.e.,
per three kagome spins, and $Z_0=\operatorname{tTr}\prod_x O(x)$, where
$\operatorname{tTr}$ sums over all internal indices.

A row of $O$ tensors forms a matrix-product operator (MPO) $\mathbb T$. We
approximate its dominant eigenvector by a uniform matrix-product state (MPS)
of bond dimension $D$, obtained with the variational uniform MPS (VUMPS)
algorithm \cite{ZaunerStauber2018,Vanderstraeten_2019,Fishman_2018}. Since
$\mathbb T$ is Hermitian, a single fixed-point problem suffices. For a row of
$M$ cells with dominant eigenvalue $\Lambda_0(M)$, the eigenvalue per cell is
\begin{equation}
 \lambda_0=\lim_{M\to\infty}[\Lambda_0(M)]^{1/M},
 \label{eq:eigenvalue-density}
\end{equation}
and the residual entropy per spin is
\begin{equation}
 s_0=\lim_{N\to\infty}\frac{\ln Z_0}{N}=\frac13\ln\lambda_0.
 \label{eq:counting-entropy}
\end{equation}

At $K/J_1=1/9$ and $D=100$ we obtain $\lambda_0\simeq4.54626264$ and
$s_0\simeq0.50476850$; the convergence in $D$ is shown in
Table~\ref{tab:ground-convergence}. This value exceeds the entropies of both
parent manifolds, $s_P\simeq0.1264$ and $s_I\simeq0.50183$
\cite{Baxter_1970,Kano_1953}, reflecting the additional mixed configurations.

\begin{table}[b]
 \centering
 \caption{Residual entropy $s_0$ per spin at $K/J_1=1/9$ versus the MPS bond
 dimension $D$.}
 \label{tab:ground-convergence}
 \setlength{\tabcolsep}{4pt}
 \begin{tabular}{lcccc}
 \toprule
 $D$ & 30 & 60 & 80 & 100\\
 \midrule
 $s_0$ & 0.504768494 & 0.504768499 & 0.504768499 & 0.504768499\\
 \bottomrule
 \end{tabular}
\end{table}

\subsection{Correlations of the equal-weight ensemble}

Having determined the ground-state degeneracy, we now consider the spin
correlations within the ground-state manifold. To this end, we calculate the spin-spin correlation function
\begin{equation}
 G_n(r)=\left\langle e^{in[\theta(r)-\theta(0)]}\right\rangle_0
       =\left\langle e^{in\pi[a(r)-a(0)]/3}\right\rangle_0,
 \label{eq:ground-correlation}
\end{equation}
where $\langle\cdots\rangle_0$ is the uniform average over the ground-state
manifold. The correlation function is evaluated by inserting impurity tensors
into the original network. The operator $e^{in\theta}$ on the spin shared
within a cell is represented by the impurity tensor
\begin{equation}
 O^{(n)}_{abcd}=\sum_g W_{abg}e^{in\pi g/3}W_{gcd}.
 \label{eq:clock-impurity}
\end{equation}
Replacing two $O$ tensors separated by $r$ cells along a row with $O^{(n)}$
and $O^{(-n)}$, and normalizing by the contraction of the original network,
gives $G_n(r)$; both contractions are evaluated with the converged boundary
MPS.

\begin{figure*}[t]
 \centering
 \includegraphics[width=0.98\textwidth]{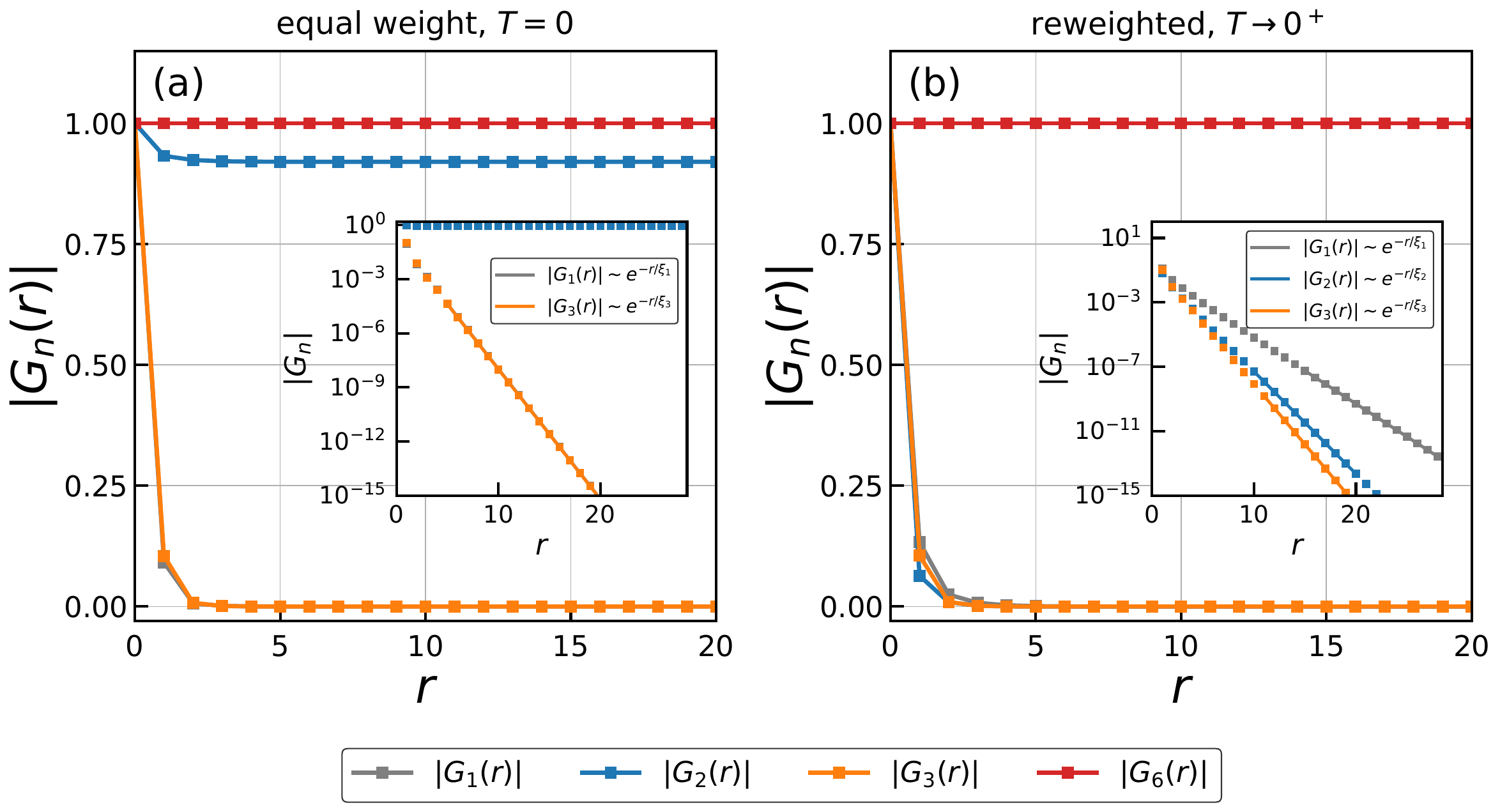}
 \caption{Spin-spin correlation functions $|G_n(r)|$ at $K/J_1=1/9$ for
 (a) the equal-weight ground-state ensemble at $T=0$ and (b) the reweighted
 ensemble at $T\to0^+$, with $D=100$. The insets show $|G_n(r)|$ for
 $n=1,2,3$ on a semilogarithmic scale together with exponential fits
 $|G_n(r)|\sim e^{-r/\xi_n}$.}
 \label{fig:ground-correlations}
\end{figure*}

Figure~\ref{fig:ground-correlations}(a) shows the result at $K/J_1=1/9$.
$G_1$ and $G_3$ decay exponentially, whereas $G_2$ saturates at
$\simeq0.920$, so the equal-weight ensemble has LRO in $e^{2i\theta}$.
This order can be understood from the mixing of the two types of ground
states: since the $I$-type manifold has a much larger residual entropy than
the $P$-type manifold, it dominates the mixed ensemble, which therefore
retains the $e^{2i\theta}$ order of the $I$-type ground states. Finally,
$G_6=1$ identically, because all allowed angles differ by multiples of
$\pi/3$.

\subsection{Fluctuation weights in the \texorpdfstring{$T\to0^+$}{T to 0+} limit}

At low temperature, thermal order by disorder must be taken into account
\cite{Villain1980Order,MoessnerChalker1998PRL,MoessnerChalker1998PRB}.
Although the ground states at $K/J_1=1/9$ have the same energy, the
fluctuations around them are not degenerate, so different ground states
acquire different statistical weights. To quantify this effect, we consider the
fluctuations around a ground state $g$. Writing
$\theta_i=\theta_i^{(g)}+\psi_i$ and expanding each bond to second order
gives
\begin{equation}
 \begin{split}
 \mathcal H&=E_0+\frac12\sum_{\langle ij\rangle}
 \kappa_{ij}(\psi_i-\psi_j)^2+O(\psi^3)\\
 &\equiv E_0+\frac12\boldsymbol\psi^{\mathsf T}
 L_g\boldsymbol\psi+O(\psi^3),
 \end{split}
 \label{eq:harmonic}
\end{equation}
where $L_g$ is the Hessian and
\begin{equation}
 \kappa_{ij}=-J_1\cos\phi_{ij}+4K\cos2\phi_{ij},
 \label{eq:kappa}
\end{equation}
with $\phi_{ij}$ the ground-state phase difference on bond $ij$.
Integrating out the fluctuations $\boldsymbol\psi$ in this harmonic
approximation gives the free energy of the ground state $g$,
\begin{equation}
 F_g=E_0+\frac T2\ln\det{}'(L_g/J_1)-c_NT,
 \label{eq:global-weight}
\end{equation}
where $\det{}'$ omits the zero mode corresponding to the global rotation and
$c_N$ is independent of the ground state. The fluctuations thus give a
correction linear in $T$, and the statistical weight of $g$ becomes
\begin{equation}
 w_g\propto e^{-F_g/T}\propto\left[\det{}'(L_g/J_1)\right]^{-1/2}.
 \label{eq:relative-weight}
\end{equation}

Evaluating the fluctuation-induced weight of an arbitrary ground state is
notoriously hard. We therefore first consider the fluctuations around the
ground states of a single triangle, and then approximate the weight by a
product of single-triangle factors, which preserves a local tensor-network
representation.

For a $P$ triangle all three bonds have $\kappa_{ij}=J_1/2-2K$, and
\begin{equation}
 L_P=\kappa_{ij}
 \begin{pmatrix}2&-1&-1\\-1&2&-1\\-1&-1&2\end{pmatrix},
 \qquad \epsilon_{P,1}=\epsilon_{P,2}=\frac32J_1-6K
 \label{eq:P-hessian}
\end{equation}
are its Hessian and nonzero eigenvalues. For an $I$ triangle with sites 1 and
2 parallel, let $a=-J_1+4K$ and $b=J_1+4K$ be the values of $\kappa_{ij}$ on the
parallel and antiparallel bonds. Then
\begin{equation}
 L_I=\begin{pmatrix}
 a+b&-a&-b\\-a&a+b&-b\\-b&-b&2b
 \end{pmatrix},
 \label{eq:I-hessian}
\end{equation}
with nonzero eigenvalues $\epsilon_{I,1}=2a+b=-J_1+12K$ and
$\epsilon_{I,2}=3b=3J_1+12K$. At $K=J_1/9$, these eigenvalues are
$\epsilon_{P,1}=\epsilon_{P,2}=5J_1/6$, $\epsilon_{I,1}=J_1/3$, and
$\epsilon_{I,2}=13J_1/3$.

The harmonic integral of a single triangle is proportional to $(\epsilon_{t,1}\epsilon_{t,2})^{-1/2}$. The relative
weight of a single $P$ and a single $I$ configuration is therefore
\begin{equation}
 \rho\equiv\frac{w_P}{w_I}
 =\sqrt{\frac{\epsilon_{I,1}\epsilon_{I,2}}
 {\epsilon_{P,1}\epsilon_{P,2}}}
 =\frac{2\sqrt{13}}5.
 \label{eq:local-ratio}
\end{equation}

\subsection{Reweighted ensemble}

Having obtained the fluctuation correction to the weight of a single
triangle, we now study its effect on the global ground states. We approximate
the weight of a global ground state by the product of the weights of its
triangles. This is implemented by modifying the local tensors of
Eq.~\eqref{eq:constraint}, so that each $P$ configuration carries the weight
$\rho$ relative to an $I$ configuration. Setting $w_I=1$, the reweighted
tensor is
\begin{equation}
 W^{(w)}_{abc}=\rho\chi_P(a,b,c)+\chi_I(a,b,c),
 \label{eq:weighted-tensor}
\end{equation}
and the reweighted partition function is
\begin{equation}
 Z_w=\sum_{\{a_i\}}\prod_t W^{(w)}_{a_tb_tc_t}.
 \label{eq:weighted-partition}
\end{equation}
The correlations are then computed as in Eq.~\eqref{eq:clock-impurity}, with
$W$ replaced by $W^{(w)}$.

As shown in Fig.~\ref{fig:ground-correlations}(b), after reweighting $G_1$,
$G_2$, and $G_3$ all decay exponentially, and the $G_2$ plateau of the
equal-weight ensemble disappears. Within this approximation, the order in
$e^{2i\theta}$ is destroyed, whereas $e^{6i\theta}$ remains locked by the six
allowed orientations of Eq.~\eqref{eq:six-state}. This suggests that
$e^{6i\theta}$ is the lowest-order field that remains coherent, and hence that
the finite-temperature phase near $K/J_1=1/9$ may exhibit QLRO in
$e^{6i\theta}$.

\section{Finite-temperature calculations}
\label{sec:finite}

\subsection{Finite-temperature tensor network}

At finite temperature we contract a tensor-network representation of the full
partition function. For frustrated spins the choice of local tensors is
important, since an unsuitable construction can impede convergence. We
therefore use the same triangle-based construction as in
Sec.~\ref{sec:ground} \cite{Song2023}, with the discrete ground-state tensor
$W_{abc}$ replaced by the Boltzmann weight of a triangle, which depends on the
continuous spin angles. To avoid large Boltzmann factors at low temperature,
we measure the triangle energy from its minimum
$h_{\min}=\min(h_P,h_I)$ and define
\begin{equation}
 W_t(\theta_1,\theta_2,\theta_3)=e^{-\beta[h_t(\theta_1,\theta_2,\theta_3)-h_{\min}]},
 \qquad \beta=1/T.
 \label{eq:boltzmann}
\end{equation}
With a normalized angular measure, the partition function is
\begin{equation}
 Z(T)=e^{-\beta(2N/3)h_{\min}}\int\prod_{i=1}^{N}\frac{d\theta_i}{2\pi}\prod_t W_t,
 \label{eq:continuous-Z}
\end{equation}
in which each angle is shared by exactly two triangle factors. We introduce an
independent angle for each triangle corner and enforce equality of paired
corners with the periodic delta function
\begin{equation}
 \delta_{2\pi}(x)=\sum_{m\in\mathbb Z}e^{imx},\qquad
 \int_0^{2\pi}\frac{d\theta}{2\pi}\delta_{2\pi}(\theta-\theta')f(\theta)
 =f(\theta').
\end{equation}
This gives the exact tensor-network representation of
Fig.~\ref{fig:construction}(b),
\begin{equation}
 \begin{split}
 Z\propto{}&\int\prod_{t,a}\frac{d\theta_{t,a}}{2\pi}
       \prod_t W_t(\theta_{t,1},\theta_{t,2},\theta_{t,3})\\
 &\times\prod_{\ell=(ta,t'b)}
       \delta_{2\pi}(\theta_{t,a}-\theta_{t',b}),
 \end{split}
 \label{eq:continuous-network}
\end{equation}
where $\ell$ runs over the links between neighboring triangles, i.e., over the
kagome spins.

\subsection{Fourier representation and transfer operator}

Since the tensors in Eq.~\eqref{eq:continuous-network} carry continuous
indices and cannot be contracted directly, we discretize them by a Fourier
transformation. With the orthonormal Fourier functions
$U_m(\theta)=e^{-im\theta}$, $m\in\mathbb Z$, the triangle tensor becomes
\begin{equation}
 I_{m_1m_2m_3}=\int\prod_{a=1}^{3}\frac{d\theta_a}{2\pi}
 W_t(\theta_1,\theta_2,\theta_3)
 e^{-i\sum_a m_a\theta_a}.
 \label{eq:fourier}
\end{equation}
Choosing conjugate Fourier conventions on up and down triangles, the basis
functions on a shared spin contract to an identity
[Fig.~\ref{fig:construction}(c)],
\begin{equation}
 \int_0^{2\pi}\frac{d\theta}{2\pi}
 U_m^*(\theta)U_{m'}(\theta)=\delta_{mm'},
 \label{eq:fourier-link}
\end{equation}
so that neighboring triangle tensors are joined directly by their Fourier
indices. In practice the Fourier indices are truncated to
\begin{equation}
 m_a\in[-n,n],\qquad d=2n+1,
 \label{eq:fourier-cutoff}
\end{equation}
where $n$ is the Fourier cutoff and $d$ is the resulting leg dimension.

Pairing an up and a down triangle gives the $O$ tensor of
Fig.~\ref{fig:construction}(d),
\begin{equation}
 O_{abcd}=\sum_{g=-n}^{n}I_{abg}I_{gcd}.
 \label{eq:finite-mpo}
\end{equation}
A row of $O$ tensors defines the transfer operator
$\mathbb T$, with dominant eigenvector
\begin{equation}
 \mathbb T|\Psi_0\rangle=\Lambda_0|\Psi_0\rangle.
 \label{eq:vumps-eigenproblem}
\end{equation}
With $\lambda_0$ the eigenvalue per cell defined as in
Eq.~\eqref{eq:eigenvalue-density}, and restoring the factor $h_{\min}$, the
free energy per spin is
\begin{equation}
 f=\frac{2h_{\min}}3-\frac{T}{3}
       \ln\lambda_0,
 \label{eq:free-energy}
\end{equation}
since a cell contains two triangles and three spins.

\subsection{VUMPS algorithm}

The dominant eigenvector of $\mathbb T$ is approximated by a uniform MPS,
\begin{equation}
 |\Psi(A)\rangle=\sum_{\{s_j\}}
 \operatorname{Tr}(\cdots A^{s_1}A^{s_2}\cdots)|\ldots s_1s_2\ldots\rangle,
 \label{eq:mps}
\end{equation}
where each $s_j$ takes $d$ values and each $A^{s_j}$ is a $D\times D$ matrix.
We obtain it using the VUMPS algorithm
\cite{ZaunerStauber2018,Vanderstraeten_2019,Fishman_2018}.
In temperature scans, the converged state at a neighboring temperature serves
as the initial state. The accuracy is controlled by $D$ and $n$, and we check
the convergence of observables in both (Sec.~\ref{sec:extra}).

\subsection{Physical quantities}

The correlation functions and thermodynamic quantities considered here can all
be expressed through expectation values of the form
$\langle e^{in(\theta_i-\theta_j)}\rangle$. We therefore first describe how an
operator $e^{in\theta_i}$ is inserted into the network. Each spin corresponds to
a link of the network, on which the two adjacent triangle tensors are joined by
the $\delta$ tensor of Eq.~\eqref{eq:continuous-network}
[Fig.~\ref{fig:construction}(b)]. Inserting $e^{in\theta_i}$ multiplies this
$\delta$ tensor by $e^{in\theta_i}$, so that after the Fourier transformation
the identity link of Eq.~\eqref{eq:fourier-link} is replaced by
\begin{equation}
 (S_n)_{mm'}=\int\frac{d\theta}{2\pi}
 e^{in\theta}U_m^*(\theta)U_{m'}(\theta)
 =\delta_{m',m+n}.
 \label{eq:shift-operator}
\end{equation}
For the spin shared inside a cell, the corresponding impurity tensor is
\begin{equation}
 O^{(n)}_{abcd}
 =\sum_{g,h}I_{abg}(S_n)_{gh}I_{hcd}.
 \label{eq:impurity-cell}
\end{equation}
An expectation value $\langle e^{in(\theta_i-\theta_j)}\rangle$ is then the
ratio between the contraction of the network with $S_n$ and $S_{-n}$ inserted
on links $i$ and $j$ and the contraction of the original network.

\emph{Correlation functions.} Replacing two cells in the same row, separated
by $r$ cells, with $O^{(n)}$ and $O^{(-n)}$ gives the spin-spin correlation
function
\begin{equation}
 G_n(r)=\left\langle e^{in[\theta(r)-\theta(0)]}\right\rangle.
 \label{eq:channel-correlation}
\end{equation}

\emph{Internal energy and specific heat.} The two spins of a nearest-neighbor
bond are legs of the same triangle tensor. Inserting $S_n$ and $S_{-n}$ on
these two legs gives the bond expectation value
$\langle e^{in(\theta_i-\theta_j)}\rangle$, and the energy of a triangle
follows as
\begin{equation}
 \langle h_t\rangle=\sum_{(ij)\in t}
 \mathrm{Re}\left[J_1\langle e^{i(\theta_i-\theta_j)}\rangle
 -K\langle e^{2i(\theta_i-\theta_j)}\rangle\right].
\end{equation}
Since there are $2N/3$ triangles, the internal energy per spin is
\begin{equation}
 u=\frac{\langle h_\Delta\rangle+\langle h_\nabla\rangle}{3}.
 \label{eq:internal-energy}
\end{equation}
The specific heat per spin,
\begin{equation}
 C_V=\frac{\partial u}{\partial T},
 \label{eq:cv}
\end{equation}
is computed from $u(T)$ by numerical differentiation along each temperature
scan.

\emph{Entanglement entropy.} Cutting the boundary MPS $|\Psi(A)\rangle$ at
one virtual bond divides it into left and right halves, with the Schmidt
decomposition
\begin{equation}
 |\Psi(A)\rangle=\sum_{\alpha=1}^{D}s_\alpha|L_\alpha\rangle|R_\alpha\rangle,
\end{equation}
where $|L_\alpha\rangle$ and $|R_\alpha\rangle$ are orthonormal states of the
two halves. The Schmidt values $s_\alpha$ are the singular values of the bond
matrix of the MPS in canonical form, which VUMPS provides directly. With
$p_\alpha=s_\alpha^2/\sum_\gamma s_\gamma^2$, the entanglement entropy is
\begin{equation}
 S_E=-\sum_{\alpha=1}^{D}p_\alpha\ln p_\alpha.
 \label{eq:entanglement}
\end{equation}
Peaks of $S_E$ as a function of temperature are used to locate the phase
boundaries.

\section{Supplementary numerical results for the phase diagram}
\label{sec:extra}

\subsection{Convergence in the Fourier cutoff}

\begin{figure}[t]
 \centering
 \includegraphics[width=0.9\columnwidth]{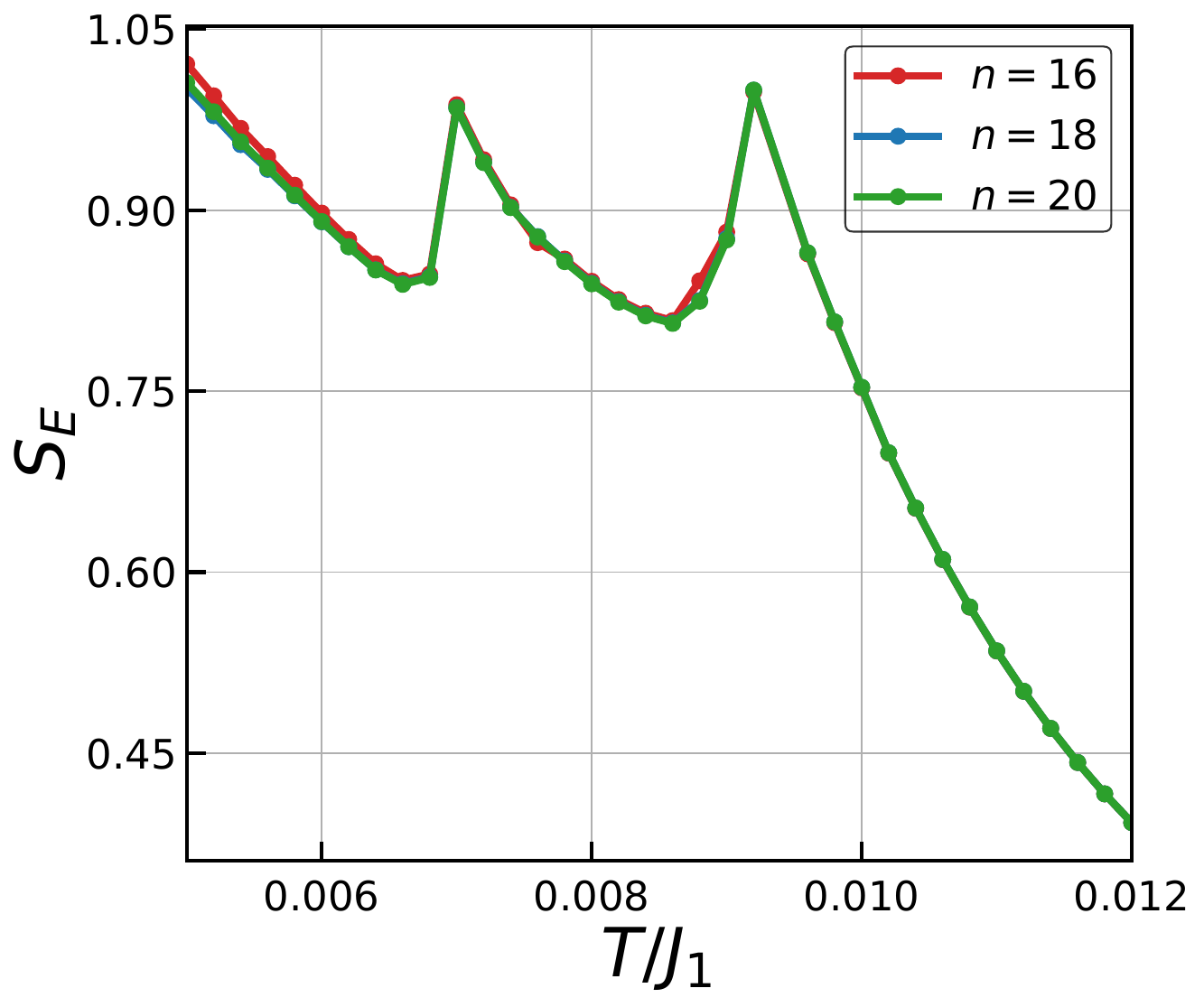}
 \caption{Entanglement entropy $S_E$ at $K/J_1=0.11$ for Fourier cutoffs
 $n=16,18,20$ at fixed $D=20$, obtained by cooling.}
 \label{fig:n-convergence}
\end{figure}

We first examine the convergence of the results with respect to the Fourier
cutoff $n$ at fixed bond dimension. Figure~\ref{fig:n-convergence} shows the
entanglement entropy $S_E$ at $K/J_1=0.11$ for $n=16$, $18$, and $20$ at
$D=20$. The three curves nearly coincide over the whole temperature range,
and the two peaks, at $T/J_1\simeq0.0070$ and $0.0092$, are located at the
same temperatures for all three cutoffs. We therefore use $n=20$ in most of
our calculations, and check the convergence at selected points by increasing
the bond dimension $D$.

\subsection{Finite-temperature behavior away from $K/J_1=1/9$}

\begin{figure*}[t]
 \centering
 \includegraphics[width=0.75\textwidth]{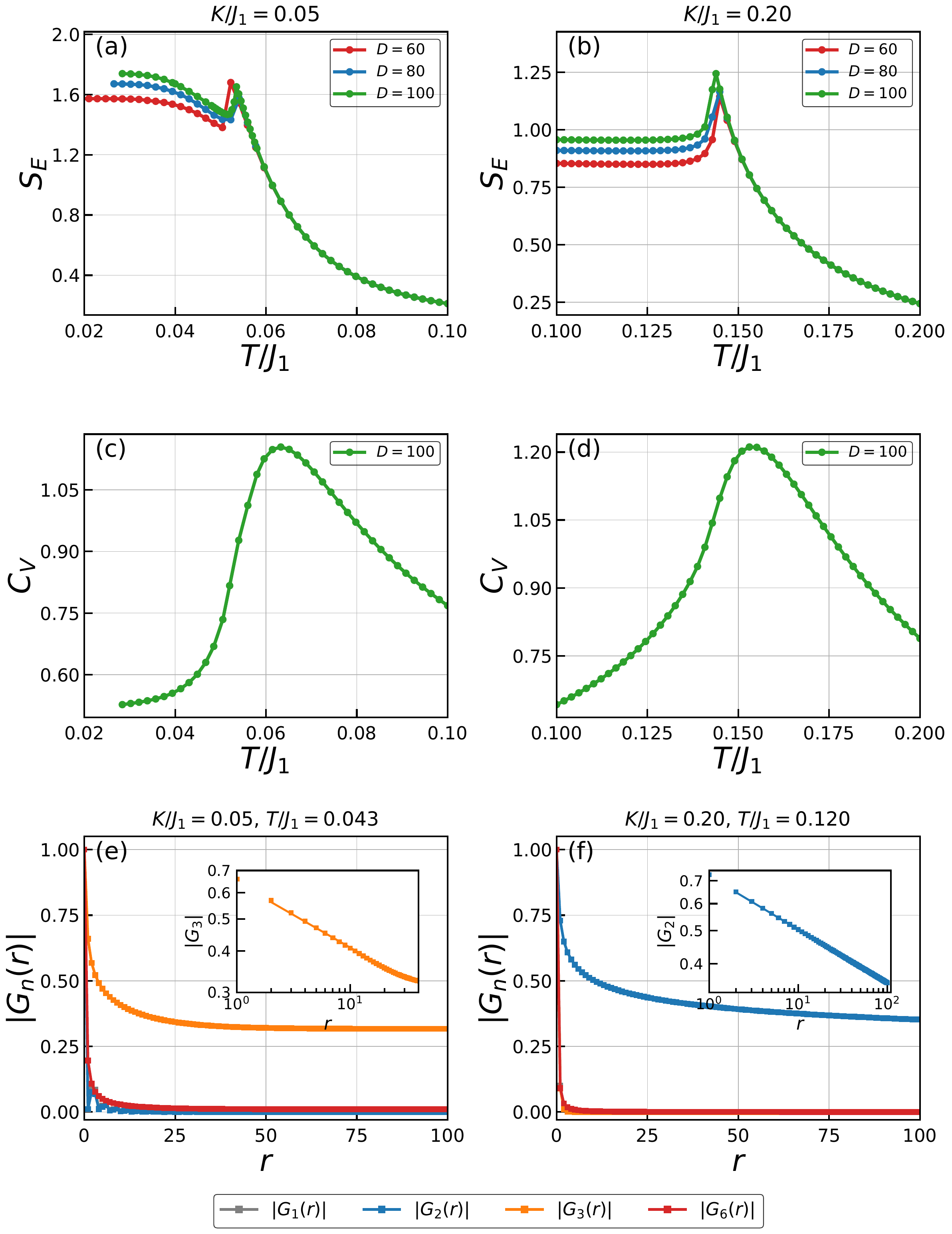}
 \caption{Behavior away from $K/J_1=1/9$ at $K/J_1=0.05$ (left)
 and $0.20$ (right), with $n=20$, obtained by cooling. (a),(b) Entanglement
 entropy $S_E$ for $D=60,80,100$. (c),(d) Specific heat $C_V$ at $D=100$.
 (e),(f) Spin-spin correlation functions $|G_n(r)|$ for $n=1,2,3,6$ in the
 low-temperature phases, at $T/J_1=0.043$ and $0.120$, respectively, with
 $D=100$. The insets show $|G_3(r)|$ and $|G_2(r)|$, respectively, on
 log-log scales together with power-law fits $|G_n(r)|\sim r^{-\eta_n}$.}
 \label{fig:away}
\end{figure*}

We now turn to the finite-temperature behavior far from $K/J_1=1/9$.
In this regime only one of the two ground-state manifolds is
energetically selected, and the intermediate $1/6$-vortex-paired phase is
absent. The system then undergoes a single transition from the
low-temperature vortex-paired phase to the disordered phase.
Figure~\ref{fig:away} shows two representative cases, $K/J_1=0.05$ on the
$P$-type side and $K/J_1=0.20$ on the $I$-type side.

In both cases the entanglement entropy exhibits a single peak, at
$T/J_1\simeq0.054$ for $K/J_1=0.05$ [Fig.~\ref{fig:away}(a)] and at
$T/J_1\simeq0.144$ for $K/J_1=0.20$ [Fig.~\ref{fig:away}(b)], which
determines the transition temperature. The specific heat shows only a broad, nonsingular maximum located
slightly above the transition temperature [Figs.~\ref{fig:away}(c) and
\ref{fig:away}(d)]. These features are
characteristic of a Kosterlitz--Thouless transition.

The nature of the low-temperature phases follows from the spin-spin
correlation functions shown in Figs.~\ref{fig:away}(e) and \ref{fig:away}(f). At
$K/J_1=0.05$, $G_3(r)$ decays algebraically [inset of
Fig.~\ref{fig:away}(e)], whereas $G_1(r)$ and $G_2(r)$ are
short-ranged. This identifies the $1/3$-vortex-paired phase inherited from
the $P$-type ground states. At $K/J_1=0.20$, $G_2(r)$ decays
algebraically [inset of Fig.~\ref{fig:away}(f)],
whereas $G_1(r)$ and $G_3(r)$ are short-ranged, identifying the
$1/2$-vortex-paired phase inherited from the $I$-type ground states. The
transitions at the entropy peaks are therefore driven by the unbinding of
$1/3$ and $1/2$ vortex--antivortex pairs, respectively. Far from
$K/J_1=1/9$, the phase diagram thus contains only the $1/3$- or
$1/2$-vortex-paired phase and the disordered phase.

\subsection{Evolution of the transitions away from $K/J_1=1/9$}

\begin{figure*}[p]
 \centering
 \includegraphics[width=0.88\textwidth]{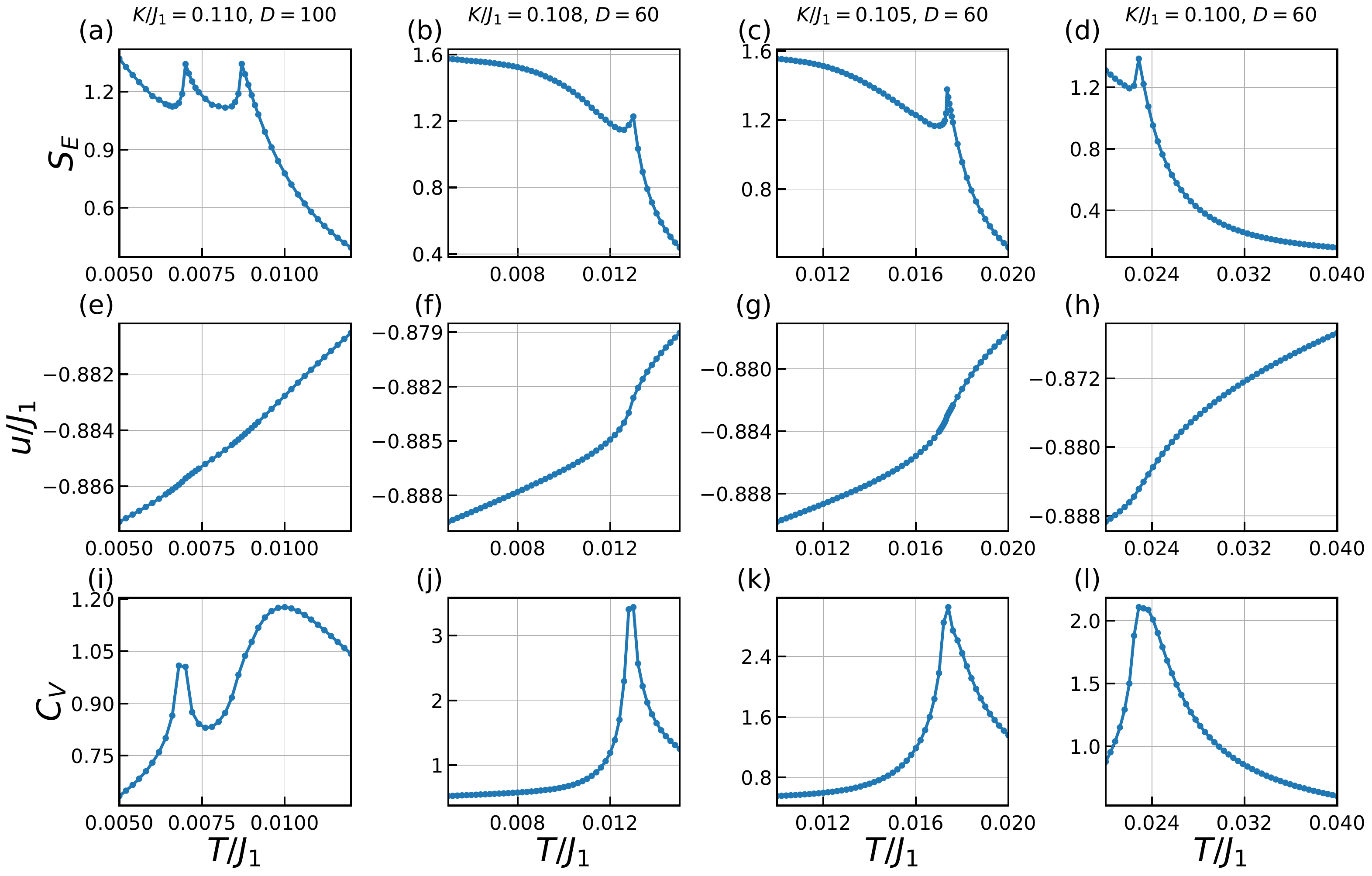}
 \caption{Evolution of the entanglement entropy $S_E$, the internal energy
 $u$, and the specific heat $C_V$ with $K/J_1$ on the $K/J_1<1/9$ side.}
 \label{fig:below}

 \vspace{12pt}
 \includegraphics[width=0.88\textwidth]{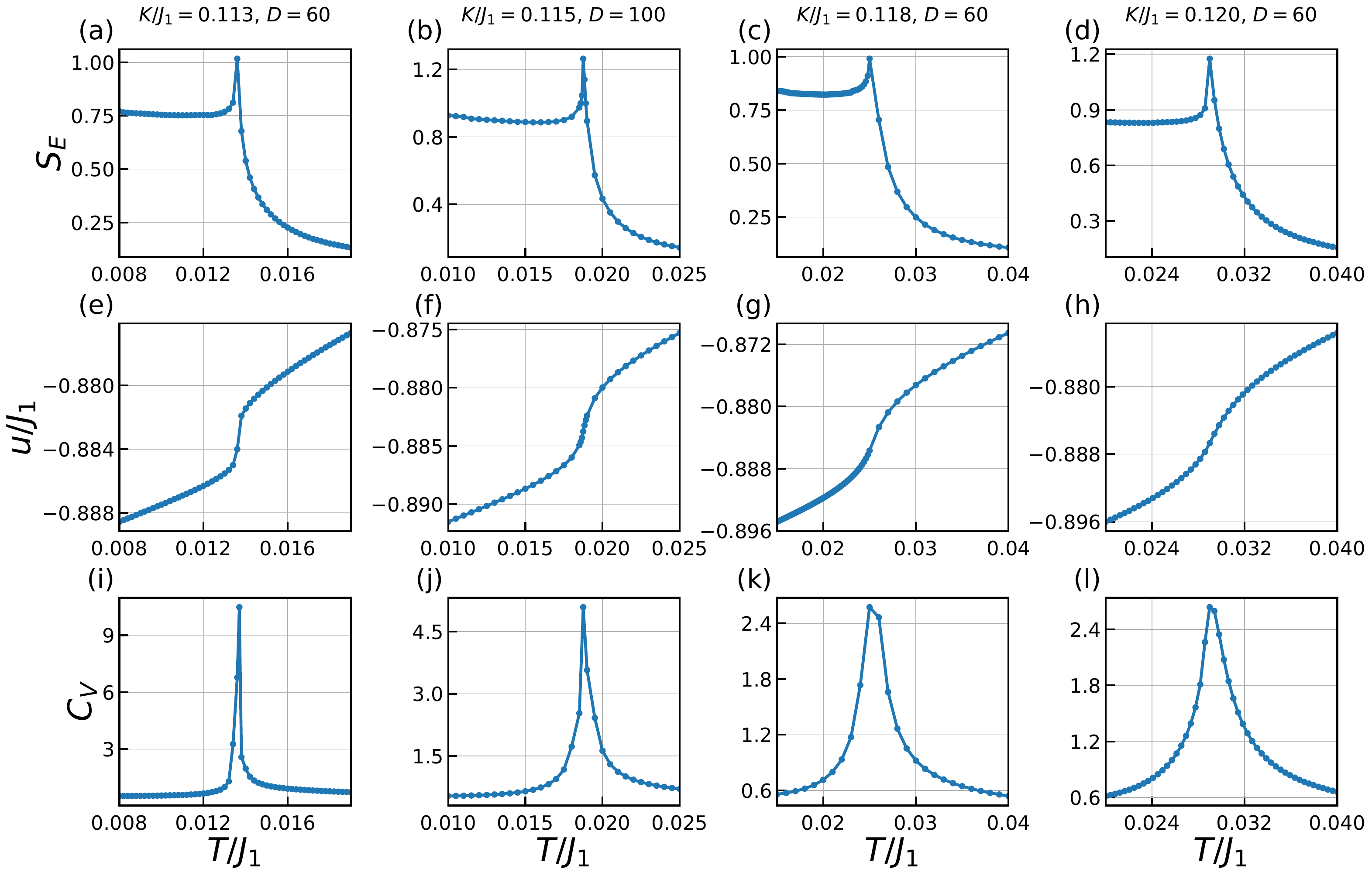}
 \caption{Evolution of the entanglement entropy $S_E$, the internal energy
 $u$, and the specific heat $C_V$ with $K/J_1$ on the $K/J_1>1/9$ side.}
 \label{fig:above}
\end{figure*}

Between $K/J_1=1/9$ and the Kosterlitz--Thouless regime discussed
above, the character of the transition from the low-temperature
vortex-paired phase to the disordered phase changes gradually with $K/J_1$.
Figures~\ref{fig:below} and \ref{fig:above} show the entanglement entropy, the
internal energy, and the specific heat for several values of $K/J_1$ on the
two sides of $K/J_1=1/9$.

On the $K/J_1<1/9$ side (Fig.~\ref{fig:below}), the two transitions that
bound the intermediate $1/6$-vortex-paired phase at $K/J_1=0.110$ merge into
a single transition at $K/J_1=0.108$. The internal energy remains smooth,
while the specific heat develops a sharp peak characteristic of a
conventional continuous transition rather than the broad maximum of a
Kosterlitz--Thouless transition, even though the system now passes directly
from the $1/3$-vortex-paired phase to the disordered phase. From the
viewpoint of topological excitations, the transition can still be understood
as being driven by the proliferation of $P$--$I$ domain walls, as for the
lower transition at $K/J_1=0.110$. Once these walls proliferate, however, the
fractional vortices bound to them are no longer paired, so that the
$1/3$-vortex-paired phase is destroyed directly and no intermediate phase
appears. As $K/J_1$ decreases further, the specific-heat peak becomes lower
and broader. Far from $K/J_1=1/9$, the $P$--$I$ domain walls cost a large
energy and become irrelevant, and the transition is driven solely by the
unbinding of $1/3$ vortex--antivortex pairs. It is then of the
Kosterlitz--Thouless type, with the broad specific-heat maximum found at
$K/J_1=0.05$ [Fig.~\ref{fig:away}(c)].

On the $K/J_1>1/9$ side, close to $K/J_1=1/9$, the system undergoes
two transitions upon heating, as shown in Figs.~\ref{fig:thermal_diagnostics}(e) and \ref{fig:thermal_diagnostics}(f) of the main
text for $K/J_1=1/9+5\times10^{-4}$. The lower-temperature transition, from
the $1/2$-vortex-paired phase to the $1/6$-vortex-paired phase, is first
order: the internal energy jumps at the transition, which produces a
divergence in the specific heat. The higher-temperature transition into the disordered
phase is of the Kosterlitz--Thouless type. As $K/J_1$ increases, the two
transitions merge,
and the resulting single transition from the $1/2$-vortex-paired phase to
the disordered phase retains its first-order character (Fig.~\ref{fig:above}):
at $K/J_1=0.113$ and $0.115$ it is accompanied by a jump in the internal
energy and a pronounced specific-heat peak. Upon further increasing $K/J_1$,
the jump is smoothed out and
the specific-heat peak is reduced, as at $K/J_1=0.118$ and $0.120$, and far
from $K/J_1=1/9$ the transition becomes of the
Kosterlitz--Thouless type [Fig.~\ref{fig:away}(d)]. In terms of topological
excitations, this evolution can be understood in the same way as on the
$K/J_1<1/9$ side: close to $K/J_1=1/9$ the transition is driven by
the proliferation of $P$--$I$ domain walls, whereas far from it these walls
become irrelevant and the transition is driven by the unbinding of $1/2$
vortex--antivortex pairs.

In summary, away from $K/J_1=1/9$ the system undergoes a single
transition on both sides, from a fractional-vortex-paired phase to the
disordered phase, but the type of this transition changes with $K/J_1$. On
the $K/J_1<1/9$ side, the transition out of the $1/3$-vortex-paired phase
evolves from a continuous transition with a sharp specific-heat peak into a
Kosterlitz--Thouless transition. On the $K/J_1>1/9$ side, the transition out
of the $1/2$-vortex-paired phase instead evolves from a first-order
transition into a Kosterlitz--Thouless transition. The mechanism underlying
this evolution, and in particular whether a regime of continuous transitions
intervenes between the first-order and Kosterlitz--Thouless regimes on the
$K/J_1>1/9$ side, remains to be explored in future work.

\end{document}